\documentclass[
 reprint,
groupedaddress,
bibnotes,
 amsmath,amssymb,
 aps,
 pra,
superscriptaddress
]{revtex4-1}
\usepackage[colorlinks,citecolor= blue,linkcolor=blue,urlcolor= blue,bookmarks=false,hypertexnames=true]{hyperref}

\usepackage{graphicx}
\usepackage{dcolumn}
\usepackage{bm}
\usepackage{ amssymb }
\usepackage[british]{babel}
\usepackage{float}
\usepackage[utf8]{inputenc}
\usepackage[T1]{fontenc}
\usepackage{textcomp}

\begin{document}

\preprint{APS/123-QED}

\title{A Comparative Study of Coupled High-Q Cavity-Quantum Dot System Regarding Dipole Induced Transparency}

\author{Alperen Tüğen}
\affiliation{
 Electrical and Electronics Engineering Department, Middle East Technical University, 06800 Ankara, Turkey
 } 
 \affiliation{
 Physics Department, Middle East Technical University, 06800 Ankara, Turkey
  \altaffiliation [Corresponding Author:] {alperen.tugen@metu.edu.tr}
 } 
\author{Serdar Kocaman}
 \affiliation{
 Electrical and Electronics Engineering Department, Middle East Technical University, 06800 Ankara, Turkey
 }

\date{\today}

\begin{abstract}
We present the differences between Input-Output formalism (IOF) and Incoherent Pumping Mechanism (IPM) derived from Lindblad Master Equation approach in terms of the transmission spectrum of Coupled high-Q Cavity with Quantum Dot system in the strong coupling regime. Full-width-half-maximum (FWHM) and the peak transmission of Dipole Induced Transparency (DIT) are inquired for detailed comparison in on-resonant and off-resonant conditions. We have found that DIT phenomenon in off-resonant case cannot be explained entirely by IPM although both methods can exhibit the same Vacuum Rabi splitting in on-resonant case. We have concluded that polariton having atomic-like feature in transmission spectrum could not be captured completely by IPM. 

\end{abstract}

\pacs{Valid PACS appear here}
\maketitle


\section{Introduction} \label{introduction}
\vspace{-0.4 cm}
Photonic Crystals (PhC) are one of the most commonly used structures in the integrated photonic devices \cite{vlasov,loncar,yablonov,gralak,fanphc,berger,kosaka}. With the latest practical advancements in cavity quantum electrodynamics (CQED), devices including PCs with quantum dots (QDs) have become important candidates for solid state CQED applications, non-classical light generation, all-optical communications and quantum logic structures \cite{maj8723,volz12}.
The most critical measure in those schemes is the level of exploitations of enhanced Purcell effect, which paves the way for strong light-matter interaction to make the nonlinear optical properties of the devices more effective \cite{vivien}.
To simulate and develop the future opto-electronical/optical integrated circuits, the studies targeting to explain the underlying mechanism that drives the dynamics of the Cavity-QD systems have been carried out in various computational methods \cite{srinivasan,kaer,gerard}. The previous studies on Cavity-QD systems have generally remodelled the expression of incoherent term in the equation of motion of the system according to the differences between those theoretical models and the observed emission/ transmission spectrum in the experiments \cite{madsen1,madsen2,calic}. On the other hand, theoretical platforms modeling these dynamical systems are still developing \cite{johansson13},
and in this paper, the comparison of two computationally favorable methods which are employed in transmission spectrum calculation of High-Q Cavity-QD systems; namely, Input-Output formalism derived from Quantum Langevin Equation \cite{gardinercollet} and Incoherent Pumping Mechanism derived from Lindblad Master Equation approach \cite{majumdaralloptical,laussy,fischer} are in focus. In particular, the transmission spectrum of the resonantly interacting high-Q Cavity-QD system has been simulated with both formalisms to analyze FWHM and the peak transmission value of the DIT (i.e. optical analogue to Electromagnetically Induced Transparency [EIT]) window in the strong coupling regime. For the case of similar spectral characters in terms of FWHM, the amplitude of the peak points of DIT \cite{sormol} calculated by IPM are significantly lower than the calculated ones by IOF. This observation shows that the features of the atomic-like polariton cannot be accurately explained by IPM. 

The plan of the manuscript is in the following; in Sec. \ref{methods}, theoretical backgrounds of Incoherent Pumping Mechanism and Input-Output Formalism is visited. In Sec. \ref{results}, detailed discussion on analysis of transmission spectra is presented. Then, the final remarks on the comparison between the methods are given in Sec. \ref{conclusion}. Moreover, for readers who are interested in further discussion on methods and the tools used for analysis, more complete information is given in Sec. \ref{appendix}. 
\vspace{-0.5 cm}
\section{Methods} \label{methods}
\subsection{Lindblad Master Equation Approach}
The corresponding state vector composition of the dissipative quantum systems is constructed by taking the average of the possible states in the ensemble. In this construction, the density matrix formalism is utilized for incorporating mixed state representation of the open quantum system into its equation of motion. In this equation of motion, while the dynamics of the  closed quantum system is governed by coherent evolution of its pure states, the dynamics regarding the interaction between closed quantum system and environment, is controlled by the random processes which take place in the environment (reservoir). Thus, to model the stochastic dissipation of the system, optical master equations which depicts both coherent and incoherent parts of the system are employed \cite{majumdaralloptical}.

The coherent part of the system is described as ordinary Heisenberg equation of motion, on the other hand incoherent part of the system is described with Lindblad form of Liouvillian super operator ( $\mathcal{L}$ ). General Liouville Equation $\dot{\rho} $ = $i[H,\rho] $ represents the unitary part of the evolution; therefore, the main aim is to extend this expression for covering the non-unitary part of the evolution that models the effect of the perturbation coming from the environment (bosonic reservoir in our case). This extension can be done by writing the evolution of density matrix with Kraus Sum representation \cite{preskill}. To obtain the Lindblad Master Equation as in the form of Eqn.\ref{eqn5}, the following approximations are necessary.
\begin{enumerate}
\setlength{\itemsep}{0pt}
\item Born approximation, which assumes initial correlation between closed quantum system and reservoir is not present due to weak coupling.
\item Markov Approximation assuming present value of  $\rho_{sys}$ has no correlation with past values.
\item Rotating Wave Approximation (RWA) (i.e. Secular Approximation), which assumes that rapidly oscillating terms in the interaction Hamiltonian average out themselves.
\end{enumerate}
As shown in the study of Karrlein and Grabert, finding a general analytical expression for Liouvillian operator is not possible for an arbitrary initial correlation between closed quantum system and environment,  since the path integral technique provides unique solutions to each possible initial preparation \cite{karrlein97}. Henceforth, the initial point of the evolution can be chosen so that no correlation between quantum system and environment is present in the beginning.
\begin{equation} \label{eqn1}
   \rho_{tot} (0) = \rho_{sys} (0) \bigotimes \rho_{E} (0)                             
\end{equation}
In experimental point of view, the initial state can be prepared to be uncorrelated with the environment or system-environment scheme can be arranged as weakly coupled.

Second approximation further considers the density function of the closed quantum system uncorrelated with the environment in later time $t_1$. 
\begin{equation} \label{eqn2}
  \rho_{tot} (t_1) = \rho_{sys} (t_1) \bigotimes \rho_{E} (t_1)                        
\end{equation}
Eqn.\ref{eqn2} means that evolution of the density matrix of overall system can be expressed in terms of direct product of density functions of the environment and the closed quantum system \cite{moy99}.With the utilization of Markov and Secular approximations alongside with the approximations mentioned(for detailed discussion about approximations see Appendix \ref{ appendix b }), the following relations can be obtained.
 In the High-Q Cavity-QD system, collapse operator corresponding to QD dipole field becomes: 
 
\begin{equation} \label{eqn3}
\dot{C_{\sigma} } = \sqrt{2\gamma}* \hat {\sigma}  
\end{equation}

where $\gamma$ is dipole spontaneous emission rate (atom dissipation rate), $\hat{\sigma}$ is the annihilation (lowering) operator of QD dipole field transition. Also, " $\dot{}$ " represents differentiation with respect to the time $t$.
The collapse operator corresponding to Cavity field is given by: 
\begin{equation} \label{eqn4}
 \dot{C_{a} } = \sqrt{2\kappa}* \hat{a}                                              
\end{equation}

where $\kappa$ is cavity field decay rate to the environment, $\hat{a}$ is the annihilation (lowering) operator of cavity field transition ($\hbar = 1$). Then, Lindblad master equation can be expressed in the following form (for detailed derivation see \cite{gardinercollet}): 
\begin{equation} \label{eqn5}
 \dot{\rho}= -i[H,\rho]+ 2\kappa\mathcal{L}(\hat{a})+2\gamma\mathcal{L}(\hat{\sigma})           
\end{equation}

where $\mathcal{L}(A)$ is Lindblad operator corresponding to collapse operator A:
\begin{equation} \label{eqn6}
\mathcal{L}(\hat{A})= \hat{A} \rho \hat{A}^{\dagger}-\frac{1}{2}\hat{A}^{\dagger}\hat{A}\rho -\frac{1}{2} \rho \hat{A}^{\dagger}\hat{A}   
\end{equation}
 
The correlation function and the spectrum of operator $\hat{A}$ of the overall system can be calculated using $Tr(\rho\hat{A}) =  <\hat{A}>$ relation, after the calculation of density matrix from the Lindblad Master Equation (Eqn.\ref{eqn5}). As mentioned above, to compare the two methods, the transmission spectrum of the Cavity-QD system is chosen, since in the steady state the Fourier transform of the two-time correlation function  corresponds to the transmission spectrum of the system when the system is weakly excited \cite{laussy,fischer}. The weak input field is introduced into equation as a Lindblad term behaving as a pump in Incoherent Pumping Mechanism discussed by Laussy et al. The Liouvilllian of the system now becomes:
\vspace{-0.1cm}
\begin{equation} \label{eqn7}
 \dot{\rho}= -i[H,\rho]+ 2\kappa\mathcal{L}(\hat{a})+2\gamma\mathcal{L}(\hat{\sigma})+2P_a\mathcal{L}(\hat{a}^{\dagger})          
\end{equation}
Normally, Wiener-Khintchine theorem \cite{wiener} states that the cavity emission spectrum can be obtained by Fourier Transform of two-time correlation function of cavity field operator, $\hat{a}$, $ S(\omega) = \lim_{t\rightarrow \infty} Re\int_{0}^{\infty}d\tau e^{i\omega\tau}<\hat{a}^{\dagger}(t+\tau)\hat{a}(t)>$. Furthermore, in the weak excitation limit, Lindblad term with $P_a$ acts as a weak incoherent photon source having broadband spectrum and the cavity emission corresponds to transmission spectrum.
\subsection{Input-Output Formalism}
Input-Output formalism is widely addressed for inspection of quantum systems whose details of subinteractions and subdynamics are not concerns \cite{gardinercollet,ekert91,fan03,gruner96}. For those cases, response of the system can be determined by relating the initial perturbation to the output \cite{ekert91}. In this formalism, the input signal is treated by damped quantum system as coupled to bosonic heat bath \cite{gardinercollet,ekert91,fan03,gruner96}. As a result of that treatment, input and output signals are represented by bath operators. Besides that, since the high frequency response of the system is mostly concerned, the dynamics between input signal and the response of the damped quantum system can support the following approximations \cite{gardinercollet}:
\begin{enumerate}
\setlength{\itemsep}{0pt}
\item System-bath interaction is assumed to be linearly related, i.e. interaction Hamiltonian must be a linear function of bath operators.
\item In the regime of operation, the coupling coefficients are assumed to be not a function of frequency .
\item Rotating Wave Approximation (Similar in Lindblad Master equation derivation).
\end{enumerate}
\raggedbottom

In addition to these approximations, since the particle is moving in a confining potential, its position does not change significantly; therefore, electric dipole approximation is applied for simplification of interaction Hamiltonian \cite{pavliotis14}. Moreover, in general, the evolution of the operators depends on the values which they take at earlier times, since the past values affect the flow of information between the environment (reservoir) and the closed quantum system. On the other hand, for the systems which have a time scale of evolution much smaller than the time scale of information exchange between the closed quantum system-environment , Markov approximation is utilized for simplifying the damping terms depending on the system operators \cite{pavliotis14,barthel12}. The motivation under these approximations is to model the system with perturbed Langevin equation, since the working regime of the open quantum system can be thought as weak external disturbance to the system in equilibrium point. Therefore, the linear response theory can be developed according to the first order perturbation introduced to the open quantum system.  Input-Output formalism developed by Gardiner et al., which is the formalism that is used here for analyzing the Cavity-QD System, also makes use of time-reversal property of input-output modes, in the same construction as quantum circuit theory \cite{yurke84}. Consequently, Heisenberg-Langevin equations (HLE) (Eqns. [\ref{eqn8}-\ref{eqn12}]) which describe the time evolution of the system are written by considering energy-conservation and time-reversal symmetry constraints \cite{fan03}. In this formalism, the same Hamiltonian is derived as the master equation approach; however, additional input-output fields, which depend on the interaction with operators of the closed quantum system and the corresponding dissipation rate into the open environment, are constructed. The following set of equations are the Quantum Langevin equations corresponding to the Coupled High-Q Cavity-QD system \cite{collet84}.

\vspace{-0.3cm}
\setlength{\belowdisplayskip}{0pt} \setlength{\belowdisplayshortskip}{4pt}
\setlength{\abovedisplayskip}{0pt} \setlength{\abovedisplayshortskip}{4pt}

\begin{equation} \label{eqn8}
\frac{d\hat{a}}{dt} = -i[\hat{a},H] - \Gamma \hat{a} +i\sqrt[]{\kappa_{1}}(\hat{a}_{in}+\hat{b}_{in})
\end{equation}
\begin{equation} \label{eqn9}
\frac{d\hat{\sigma}}{dt} = -i[\hat{\sigma},H] - \gamma \hat{\sigma}
\end{equation}
\begin{equation} \label{eqn10}
\hat{b}_{out} = \hat{a}_{in} + \sqrt[]{\kappa_{1}}\hat{a}
\end{equation}
\begin{equation} \label{eqn11}
\hat{a}_{out} = \hat{b}_{in} + \sqrt[]{\kappa_{1}}\hat{a}
\end{equation}    
\begin{equation} \label{eqn12}
H = \omega_{c}\hat{a}^{\dagger}\hat{a} + \omega_{r}\hat{\sigma}^{\dagger}\hat{\sigma} + [-ig(\vec{r})\hat{\sigma}^{\dagger}\hat{a}+h.c.]
\end{equation}
\vspace{-0.2 cm}

\noindent where $2\Gamma$ is the total cavity decay rate, where $\Gamma =\frac{\kappa_0+\kappa_1}{2}$, $\kappa_0$ is the intrinsic cavity decay rate, $\kappa_1$ is external cavity decay rate, $g(\vec{r})$ is the coupling strength between the cavity mode and QD mode \cite{njparticle}.Then, the transmission matrix can be found by dividing output field by input field matrix.

\section{Results} \label{results}
The simulations have been performed under the approximations stated in Sec.\ref{methods}. At this stage, it is critical to note that the system parameters, $\Gamma$ and $\gamma$, which are defined to have same values for the sake of consistency in the calculations.(For further details about simulations, please refer Appendix \ref{ appendix a }). Moreover, it is pertinent to note that, in this study, the corrections concerning the linewidths of the polaritons have not been included, since those corrections do not change the characteristic shape significantly \cite{majumdarline,majumdaralloptical} and two methods do not inherently have any of these corrections in their original form. Corollary, these two methods can be compared without loosing generality.

\begin{figure}[H] 
\centering
\includegraphics[width=\linewidth]{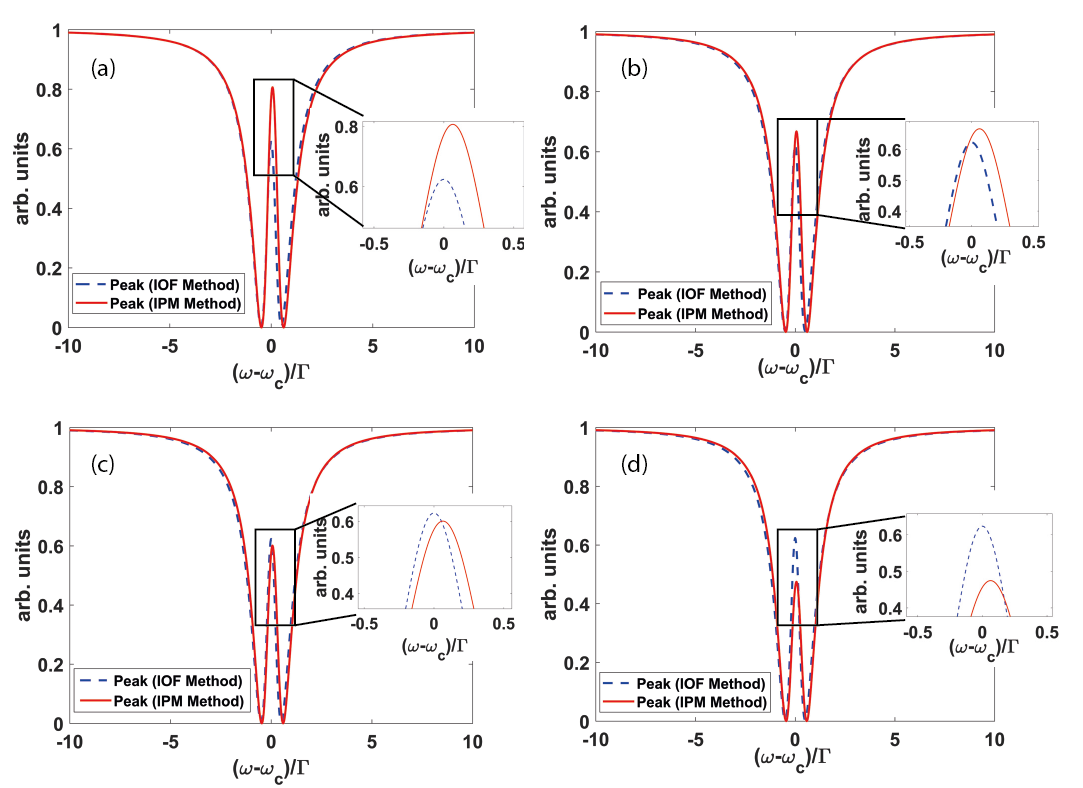}
\caption{Transmission spectrum comparison between Input-Output formalism and Incoherent Pumping Mechanism utilizing Lindblad Master Equation for $P_a = 1,2,2.5$ and $3.5\gamma$,in a,b,c,d respectively when $\Gamma = 15\gamma$.}
\label{fig:fig1}
\end{figure}

\begin{figure}[H] 
\centering
\includegraphics[width=\linewidth]{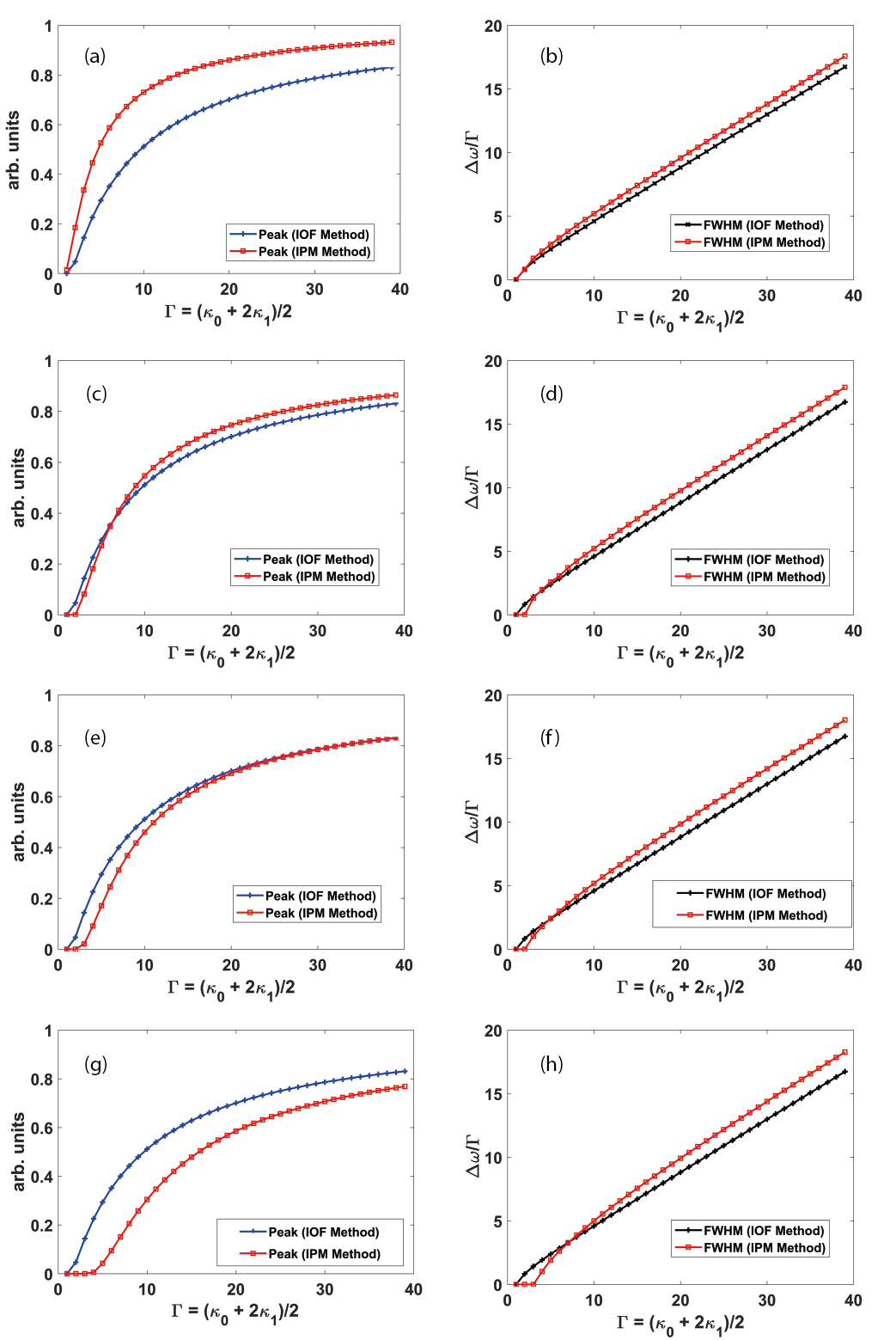}
\caption{Peak and Full-Width-Half-Maximum of Dipole Induced Transparency (DIT) comparison between Input-Output formalism and Incoherent Pumping Mechanism utilizing Lindblad Master Equation for $P_a = 1,2,2.5$ and $3.5\gamma$, as (Peak,FWHM) in (a,b), (c,d), (e,f), (g,h), respectively.}
\label{fig:fig2}
\end{figure}

In simulations, the decay rate, $ \kappa $ value have been swept, while the coupling strength, $g$, is kept equal to $\frac{\kappa}{2}$. Thus, $g^2 > \frac{(\kappa - \gamma)^2}{16}$ condition is satisfied which ensures the system operates in the strong coupling regime \cite{reithmayer}. As the comparison has been tried to draw between rates equations of the methods, it is salient that the $P_a$ term, which is incorporated in  Incoherent Pumping Mechanism, does not coincide with any term in rate equations of Input-Output Formalism. Since $P_a$ has no correspondence, to find the exact fit for the peak and FWHM of DIT, $P_a$ values have been swept. Fig.\ref{fig:fig1} shows four transmission spectra obtained with $P_a$ values changing from $1$ to $3.5$ in on-resonant case, i.e. $\omega_{cav} = \omega_{QD}$. The overlapping between the transmission spectra is apparent in Fig. \ref{fig:fig1} (c). In the further quantitative analysis, the FWHM and peak values of DIT has been investigated by sweeping decay rate, $\Gamma$, in Fig. \ref{fig:fig2}. The match is found for $P_a$ = 2.5 $\gamma$. 
In the graphs, the agreement between the methods in peak values becomes more explicit after the near-bad-cavity limit ($\gamma << \kappa$) \cite{fischer,rice88} on-resonant case. The reason that we have obtained a better fit for large $\Gamma$ values can be explained by inquiring the components of the transmission spectrum of the system. Transmission spectrum can be decomposed into two stationary spectra of two output channels, namely side and axis emissions. Side emissions of system are characterized by the autocorrelation of the atomic annihilation/ creation operator, i.e. $ T_{side}(\omega) = \frac{\gamma}{2\pi}\int_{0}^{\infty}d\tau e^{-i\omega \tau}<\sigma^{\dagger}(t+\tau)\sigma(t)>$, while axis emissions are characterized by cavity field annihilation/ creation operator, i.e. $T_{axis}(\omega) = \frac{\kappa}{\pi}\int_{0}^{\infty} d\tau e^{-i\omega \tau} <a^{\dagger}(t+\tau)a(t)>$. These spectra are related to each other through normalization in an way that $\int_{-\infty}^{\infty}d\omega T_{side}(\omega) +\int_{-\infty}^{\infty}d\omega T_{axis}(\omega) = 1 $ \cite{carmichael}. 

\vspace{-0.2cm}

In the calculations, the transmission spectrum is approximated to only consist of cavity emission ($T(\omega)$ = $\propto$ $ <\hat{a}^{\dagger}\hat{a}>$) and contribution coming from the direct coupling between atomic emission to free continuum modes at outside of Cavity-QD system is neglected, since the following two reasons. First, the nonlinear relation between $T_{axis}$ and $T_{side}$ makes the calculations intractable in the computational point of view; second, the ratio between $\gamma$ and $\Gamma$ makes $T_{side}(\omega)$ insignificant with respect to $T_{axis}(\omega)$ \cite{fischer,laussy,reithmayer}. Hence, the matching between the peaks is anticipated when the system is in the verge of bad-cavity limit. In the light of this analysis, determining the fitting parameter $P_a$ by considering peak matching for values of $\Gamma$ satisfying near-bad-cavity condition, indeed makes sense.

However, even though the agreement in the peak value can be set, the perfect matching for FWHMs has not been obtained. When Fig.\ref {fig:fig2} (b) is inspected, the linear relation between FWHMs and $\Gamma$ is apparent albeit small difference between FWHM values is present in near-good cavity limit. On the other hand, for increasing value of the $P_a$, a shift is observed in the first $\Gamma$ value, at which transmission spectrum of the system has started to form a peak at resonant frequency. In other words, IPM has failed to explain peak formation at resonant frequency for near-good-cavity limit, which can be seen in first two, two and three red square dots in Fig.\ref{fig:fig2} (d,f,h) respectively. That failure introduces a discrepancy in FWHM values which is changing between $6.7\%-10\%$. Nonetheless, for the large values of $\Gamma$ (near-bad-cavity limit), the linewidth become linearly related with $\Gamma$, which can be seen in the trend of black (plus) and red (square) lines in Fig.\ref{fig:fig2} (b,d,f,h). As more detailed numerical analysis is performed on the linewidth graphs, the linear relation between linewidth and $P_a$ is observed. That is an expected result since the previous studies showed a linear relation between linewidth and decaying term, $\gamma$ \cite{majumdarline,majumdarthesis} which is introduced to the master equation in the same way as $P_a$. Therefore, in experimental point of view, two methods converge same $\Gamma $ and $\gamma$ values in bad-cavity limit.

\begin{figure}[H] 
\centering
\includegraphics[width=\linewidth]{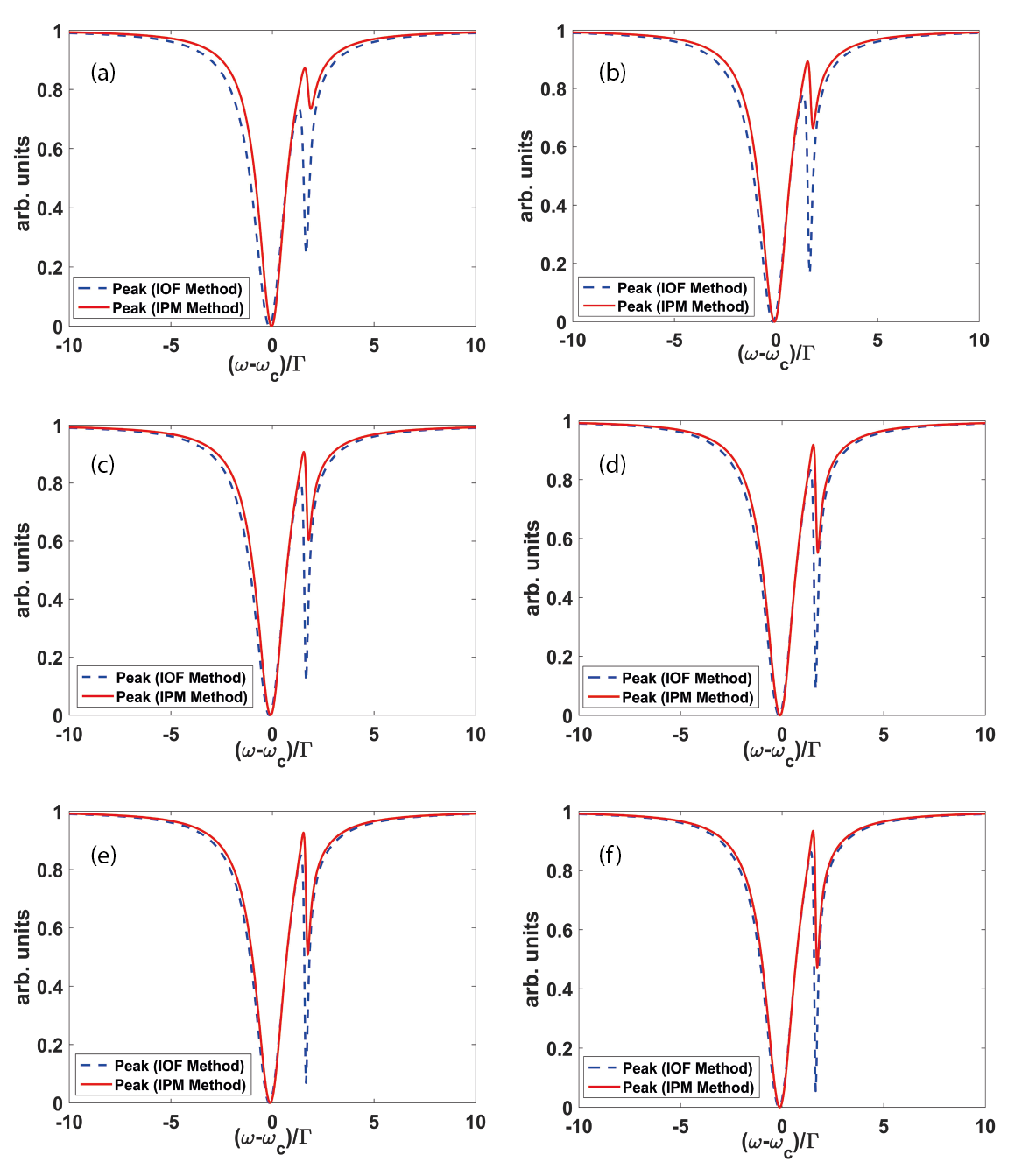}
\caption{Transmission Spectrum of High-Q Cavity-Quantum Dot system with $\omega_{QD}-\omega_{cav} = 1.5\Gamma$ detuning obtained by Input-Output Formalism and Incoherent Pumping Mechanism method utilizing Lindblad Master Equation, $\Gamma = 10,15,20,25,30$ and $35 \gamma$, in a, b, c, d, e ,f respectively.}
\label{fig:fig3}
\end{figure}

To complete the discussion on transmission spectrum of Cavity – QD system, and articulate the effect of atomic emission on the transmission spectrum, the system has been investigated in off-resonant case as well and the results are summarized in Fig.\ref{fig:fig3}. The detuning between cavity and QD frequency has been chosen so that transmission spectrum have both cavity-like and atom-like polaritons distinctly and, at the same time, the effect of Dipole Induced Transparency (DIT) phenomenon is present in the spectrum. 
This configuration can be easily obtained for $\delta$ value which is sufficiently greater than $g$. Since the linewidth of the cavity and the atomic emission is approximately given by $2\Gamma + 2 (\frac{g}{\delta})^2\gamma $ and $2\gamma + 2 (\frac{g}{\delta})^2\Gamma $, respectively \cite{majumdarline}.
Accordingly, the Cavity-QD system has been put into a $ \delta = 1.5 \Gamma$ detuned regime, and the left polariton is assured of having more cavity-like characteristic, while the right polariton has atomic-like characteristics.

Also, in the calculations, atomic dephasing mechanism has been excluded  since dephasing might compensate the effect of cavity-QD detuning undesirably \cite{auffeves}. In addition, this feature of the system dynamics is not incorporated within the Input–Output Formalism. Hence, for meaningful comparison between methods this mechanism is disregarded as together with the discussion about the additional corrections mentioned in the beginning of this section. 

As a result, in the analysis of Cavity-QD System in off-resonant case, in a similar reasoning as the on-resonant case, Incoherent Pumping Mechanism method expectedly lose the signature of the Dipole Induced Transparency peak, as can be seen in Fig.\ref {fig:fig3} (a,b), whereas the effect of atomic emission on the transmission spectrum is observed in Input-Output Formalism. This discrepancy is, again, resulted from the fact that the magnitude of the right dip is mostly related with the contribution coming from atomic emission. In addition, the considerable variation in the linewidth is observed in the absence of incoherent coupling \cite{majumdarline} of atomic emission to the cavity emission. The absence of $T_{side}$ emission creates noticeable difference on FWHM of left (cavity-like) polariton, shown as blue (dashed) and red (solid) curves in Fig.\ref{fig:fig3} (a,b). After this anticipated discrepancy in near-good-cavity limit regime is observed, the system parameters have been changed so that the near-bad-cavity operation is facilitated for testing the performance of methods on explaining transmission spectrum in off-resonant case. As the system goes beyond bad-cavity limit, both cavity linedwidth values gradually come to an agreement, as depicted in Fig.\ref{fig:fig3} (c-f).

Nonetheless, more strikingly, the magnitude of the dips are not matched although the near-bad-cavity condition is satisfied in off-resonant case, as in Fig \ref{fig:fig3} (e,f). Besides that, the peak values of Dipole Induced Transparency are not matched in both good and bad cavity limit, even though rough agreement in the peaks become more noticeable through near-bad-cavity limit.
These results indicate that Incoherent Pumping Mechanism approach cannot fully describe the peak of the polariton which has atom-like characteristic by just simply calculating autocorrelation function of cavity field operator. Because the experiments show the value that emission spectrum takes at $\omega_{cav}$, which is nearly equal the value that emission spectrum takes at $\omega_{QD}$. On the contrary, the dip value of the right polariton, calculated by Incoherent Pumping Mechanism approach, is smaller between $ 50\% - 62.5\% $ than the value estimated by Input-Output formalism, whose results show strong agreement with the emission spectrum experiments \cite{majumdarline,hennessy,yoshie}.
\section{Conclusion} \label{conclusion}
The methods that we have discussed in this paper, are currently used for estimating the parameters of the system alongside with  experimental results for various open quantum system schemes, especially for Cavity-Multi QDs systems. Therefore, testing their capability of explaining the experimental results is crucial. Consequently, the differences between theory and experiments can help to refine the approximations or help to refine the theory all together. Here, we have observed that even though the approximations utilized by Input-Output formalism and Incoherent Pumping Mechanism method are overlapping in the strong coupling regime for High-Q Cavity-QD systems, significant  discrepancies on the DIT transmission peak characteristics in off-resonant case have been observed. In addition to that result, small differences in FWHM values of DIT both on and off-resonant case have been detected. However, in order to keep the comparison as simple as possible with including all the important dynamics, we have not included the corrections for linewidth of the polaritons reported in other studies \cite{majumdarline,majumdaralloptical}.
As a final remark, further computational advancements in calculations of transmission and emission spectrum, will lead more easy and reliable comparison between theoretical and experimental studies. Thus, the understanding of the competence of various computational methods on explaining physical dynamics of Quantum systems, and the possible corrections \cite{kocaman16,majumdarline,majumdaralloptical} to these models, will remain an integral part for more accurate future all-optical device designs.
 \section{Appendices} \label{appendix}

 \subsection{Correspondence of System Parameters} \label{ appendix a }
 To facilitate the correspondence between Input-Output Formalism and Incoherent Pumping Mechanism, we have shown the parameters $\kappa$ and $\gamma$ are same for both methods by comparing rate equations of cavity and atom operators of each method.As expressed before, the dynamics of the Cavity-QD System governed by the following equation where $\rho_I$ and $H_I$ density function and Hamiltonian in interaction picture :
 \begin{equation} \label{eqna1}
 \dot{\rho_I} = -i[H_I,\rho_I] + 2\kappa\mathcal{L}(a) + 2\gamma\mathcal{L}(\sigma) +2P_a\mathcal{L}(a^{\dagger}) 
 \end{equation}
 Also noting that $<\dot{A}_I> = Tr(A_I\dot{\rho}_I)$ where $\hat{A}_I$ is any operator in interaction picture. Hence, we can obtain following relation by using Eqn.\ref{eqna1} and cyclic property of Trace operation:

 \begin{equation} \label{eqna2}
 \begin{split}
 <\dot{A}_I> = -i<[A_I,H_I]> + 2\kappa<\mathcal{L'}(a)> + 2\gamma<\mathcal{L'}(\sigma)> \\ +2P_a<\mathcal{L'}(a^{\dagger})> 
 \end{split}
 \end{equation}
where $\mathcal{L'}(D) = D^{\dagger}A_ID- \frac{1}{2}D^{\dagger}DA_I - \frac{1}{2}A_ID^{\dagger}D $. One can obtain the same rate equations apart from input and output field operators and Langevin Noise terms if $a$ and $\sigma$ operators are inserted into Eqn.\ref{eqna2}. In other words, cavity and QD decay rate are expressed with same dissipation rate parameters for both methods.

 Furthermore, $\lim_{t\rightarrow \infty}<a^{\dagger}a>(t)$ is calculated by solving 5 coupled differential equations including $<\sigma^{\dagger}\sigma>$, $<a^{\dagger}\sigma>$, 
 $<\sigma^{\dagger}a>$ by empowering Eqn\ref{eqna2}. Those differential equations are converted to be linear equations when their steady state values are the main interest since $\lim_{t\rightarrow \infty}\frac{d}{dt}<a^{\dagger}a>(t)= 0$. As a result of this calculation, cavity population ($n_a$) on resonant case is obtained as:
 \begin{equation} \label{eqna3}
 n_a = \frac{P_a(g^2+\chi\gamma)}{\chi(\xi\gamma+g^2)} 
 \end{equation}
 where $\chi = \xi + \gamma$ and $\xi = \kappa - P_a$. This result can easily be generalized to off-resonant case by changing interaction Hamiltonian \cite{laussy}.

 \subsection{Further Discussion on Approximations} \label{ appendix b }
As mentioned before, the equations describing open quantum systems obtained from Quantum Stochastical Differential Equations are usually consisting of intractable integrals. Hence, reasonable and justifiable assumptions are necessary for calculating the observables of the system computationally. Subsequently, approximations have been made in Section 2, are elucidated more. We have assumed that evolution of the density matrix corresponding our total system can be considered as uncorrelated with environmental degrees of freedom, as seen Eqn.\ref{eqna4}.

\begin{equation} \label{eqna4}
  \rho_{tot} (t_1) = \rho_{sys} (t_1) \bigotimes \rho_{E} (t_1)                        
\end{equation}

This approximation is too important for obtaining more manageable equation of evolution. On the other hand, at first glance, this approximation looks contradictory to physical intuition on evolution of the system which results in an internal system becomes more entangled with the reservoir, and this entanglement is the reason for the pure state evolves into a mixed state. Therefore, the correlation between individual density matrix of system and environment increases. In order to justify the assumption, two features of the system-environment interaction should be addressed \cite{breuer}. First, the entanglements between system and degrees of freedom of the environment cannot be tracked easily, and this makes redundant to assign or construct a corresponding state to the environment. Second, the information leaking out of the system to the environment is not likely to come back from the environment to the system, at least when the overall system made up of Electromagnetic field, and furthermore this type of open system realizations do not show repetition of "talk" between internal system and environment. Overall, these two features enable us to approximate the system having almost uncorrelated dynamic.Thus we are able to obtain integral expression in Eqn.\ref{eqna5} \cite{breuer}:

\begin{equation} \label{eqna5}
  \frac{d}{dt}\rho_{sys} (t) = -\int_{0}^{t} ds tr_E [H_I(t),[H_I(s), \rho_{sys} (t_1) \bigotimes \rho_{E} (0)]]                        
\end{equation}

Third important approximation is Markov approximation which indicates the correlation function of the environment decays rapidly compared to time scale of the evolution of the whole system. Therefore, under this approximation system is not affected too much by the past values of the density matrix of the Coupled High-Q Cavity-QD system \cite{gerry2005}. The physical justification of the Markov approximation is highly dependent on the spectrum of the bath, in our system bosonic harmonic oscillators in the environment \cite{davies1980}. The final approximation is Secular approximation, or Rotating Wave approximation \cite{breuer}. This approximation is used for simplifying the interaction Hamiltonian by neglecting the terms making much faster transition than the time scale of the evolution by considering that the time average of those terms is rapidly goes to 0 \cite{johansson13}. In addition , with further expansion of the Kraus operator, familiar form of the Lindblad Master Equation including non-unitary evolution terms can be obtained.

\end{document}